\documentclass{Interspeech}

\interspeechcameraready

\title{Synthetic Speech Source Tracing using Metric Learning}

\author[affiliation={1,2}]{Dimitrios}{Koutsianos}
\author[affiliation={1}]{Stavros}{Zacharopoulos}
\author[affiliation={2,3}]{Yannis}{Panagakis}
\author[affiliation={1,2,4}]{Themos}{Stafylakis}

\affiliation{Department of Informatics}{Athens University of Economics and Business}{Greece}
\affiliation{}{Archimedes/Athena RC}{Greece}
\affiliation{Department of Informatics and
Telecommunications}{National and Kapodistrian University of Athens}{Greece}
\affiliation{}{Omilia - Conversational Intelligence}{Greece}
\email{dkoutsianos@aueb.gr, yannisp@di.uoa.gr, tstafylakis@aueb.gr}

\usepackage{comment}
\usepackage{mathrsfs}
\usepackage{multirow}
\usepackage{float}

\begin{document}

\maketitle

\begin{abstract}

This paper addresses source tracing in synthetic speech—identifying generative systems behind manipulated audio via speaker recognition-inspired pipelines. While prior work focuses on spoofing detection, source tracing lacks robust solutions. We evaluate two approaches: classification-based and metric-learning. We tested our methods on the MLAADv5 benchmark using ResNet and self-supervised learning (SSL) backbones. The results show that ResNet achieves competitive performance with the metric learning approach, matching and even exceeding SSL-based systems. Our work demonstrates ResNet's viability for source tracing while underscoring the need to optimize SSL representations for this task. Our work bridges speaker recognition methodologies with audio forensic challenges, offering new directions for combating synthetic media manipulation.
\end{abstract}

\section{Introduction}

Recent advances in audio manipulation, particularly text-to-speech (TTS) models \cite{casanova24_interspeech, pmlr-v139-kim21f,NEURIPS2020_5c3b99e8} have enabled synthetic media to be indistinguishable from human speech. However, these innovations have also been exploited for malicious purposes \cite{hutiri2024not}, driving the need for robust anti-spoofing systems. Although detecting spoofed audio remains paramount, identifying the source system responsible for generating synthetic content offers critical forensic value, aiding authorities in tracing malicious actors.

Existing approaches to source tracing focus on attribute classification. The authors of \cite{klein24_interspeech} proposed end-to-end and two-stage classifiers to jointly identify input types, acoustic models, and vocoders in spoofed audio. Others target specific components, such as vocoders \cite{10.1145/3552466.3556525} or attack systems in datasets like ASVSpoof2019 \cite{DBLP:journals/ejisec/BorrelliBAST21}. In parallel, speaker recognition addresses training either as closed or as an open-set identification task. In closed settings, a classification head is added to the embedding extractor and loss functions such as A-Softmax \cite{8100196}, AM-Softmax \cite{8578650}, and AAM-Softmax \cite{8953658} which help the extracted embeddings generalize to new speakers during testing. In open settings, framed as metric learning problems, embedding extractors are trained without a classification head, using loss functions
such as Generalised end-to-end (GE2E) \cite{10.1109/ICASSP.2018.8462665}, triplet loss \cite{schroff2015facenet}, and prototypical loss \cite{10.5555/3294996.3295163}. 

The authors of \cite{qin2023speaker} address a similar problem to ours, although with all but one TTS system in testing also included in training. They implemented a center-based maximum similarity method by creating simple centroids for each class and after extracting the audio embeddings of a fusion model they calculated the cosine similarity between each audio embedding and each centroid, if the similarity was larger than a threshold they assigned it to the class with the larger similarity, else they assigned it to an unknown 7th class.

Building on \cite{chung20b_interspeech}, who compared loss functions for speaker verification and introduced Angular Prototypical loss, we investigate source system identification using two architectures: a Thin-ResNet \cite{chung2019delving} and a SSL model integrating AASIST \cite{jung2022aasist} with fine-tuned Wav2Vec2 embeddings \cite{baevski2020wav2vec}. 

We evaluate both architectures with different embedding dimensions, loss types (classification, metric learning) and sampling strategies (balanced vs. random). Performance is assessed with Equal Error Rate (EER) using the cosine similarity between two audio embeddings.

Our results demonstrate that Thin-ResNet achieves a performance comparable to that of the SSL model, despite significantly lower training complexity, positioning it as a practical solution for source tracing.

\section{Losses and metric learning methods}

We now present the methods tested during our research, categorized into two groups: classification loss functions and metric learning methods.

\subsection{Classification Loss Functions}

The training set of the MLAADv5 \cite{10650962} dataset, and more specifically a version for source tracing\footnote{\url{https://deepfake-total.com/sourcetracing}}, contains $\mathcal{T}=24$ distinct TTS systems. During training, whether the mini-batch was balanced across the TTS systems or not, each mini-batch contains $N$ distinct audio samples, with $C$-dimensional embeddings $\mathtt{x}_i\in \mathbb{R}^{C}$, $i\in [N]$, and respective labels $\mathtt{y}_i$, $i \in [N]$. 

\subsubsection{Softmax}

Softmax loss is nothing short of a softmax activation function followed by a Cross-Entropy Loss:
\begin{equation}
\mathcal{L}_s = -\frac{1}{N}\sum_{i=1}^N \log \frac{\exp({W_{\mathtt{y}_i}^T \mathtt{x}_i + b_{\mathtt{y}_i}})}{\sum_{j} \exp({W_j^T \mathtt{x}_i + b_j})},\, j \in [\mathcal{T}]
\end{equation}
where we denote $W$ and $b$ the weight and bias of the last layer of the model (i.e. the classification head).

\subsubsection{AMSoftmax}

The AMSoftmax loss is derived from the standard Softmax loss where instead of the dot product between the embeddings and the weights of the embedding layer, it uses the cosine of the normalized dot product of the embeddings and the weights of the embedding layer. A margin $m$ is added to the cosine of the target class, which encourages better discrimination and generalization to new classes during testing.

\begin{equation}
\mathcal{L}_a = -\frac{1}{N}\sum_{i=1}^N \log \frac{\exp({s(\cos(\theta_{\mathtt{y}_i,i}) - m)})}{\sum_{j} \exp({s(\cos(\theta_{j,i})-m\delta_{\mathtt{y}_i,j})})}    
\end{equation}
where $\delta_{\cdot,\cdot}$ is the Kronecker delta function and $s>1$ a scaling factor of the logits.

\subsubsection{AAMSoftmax}

The AAMSoftmax loss is similar to the AMSoftmax loss, but instead of subtracting the margin from the cosine, it adds it to the angle:

\begin{equation}
    \mathcal{L}_A = -\frac{1}{N}\sum_{i=1}^N \log \frac{\exp({s\cos(\theta_{\mathtt{y}_i,i} + m))}}{ \sum_{j} \exp({s\cos(\theta_{j,i}+m\delta_{\mathtt{y}_i,j}}))}
\end{equation}

\subsection{Metric Learning Methods}

For the metric learning methods we utilized the balanced version of the mini-batches. Therefore, each mini-batch contains exactly $M$ distinct audio samples, balanced across $N$ different TTS systems. This means that $M = \kappa N$ where $\kappa$ is the number of distinct audio samples from the same class that are featured in each mini-batch.

For both methods that we mention in this section, we denote by $S$ the support set of each batch, which consists of all the utterances of the batch, and by $Q$ the query set of each batch, which consists of only one utterance per TTS model. For simplicity, assume that the query set is the $M^\textit{th}$ utterance of each model. 

\subsubsection{Generalized End-to-End}

In GE2E, during training, centroids are computed using all utterances in a batch except for the query itself. Specifically, for each class, the centroid is calculated from all available utterances, while the centroid corresponding to the query's class is computed using one fewer utterance than those of other classes. These centroids are defined as follows:

\begin{equation}
    c_j = \frac{1}{M}\sum_{m=1}^M \mathtt{x}_{j,m}, \,\, c_j^{(-i)} = \frac{1}{M-1}\sum_{\substack{m=1\\ m\neq i}}^M \mathtt{x}_{j,m} 
\end{equation}

The similarity matrix is computed using scaled cosine similarity between the embeddings and all centroids:

\begin{equation}
 \mathcal{S}_{i,j,k} =
  \begin{cases}
  w\cdot \cos(\mathtt{x}_{j,i},c_j^{(-i)}) + b & \text{if } k=j \\
  w\cdot \cos(\mathtt{x}_{j,i},c_k) + b & \text{otherwise}
  \end{cases}
\end{equation}

where $w > 0$ and $b$ are learnable scaling and bias parameters. The final loss is given by

\begin{equation}
    \mathcal{L}_g = -\frac{1}{N} \sum_{j,i} \log \frac{\exp{\mathcal{S}_{j,i,j}}}{\sum_{k=1}^N \exp{\mathcal{S}_{j,i,k}}}
\end{equation}

\subsubsection{Angular Prototypical}

An extension of the standard prototypical loss, where, during training, each query is classified against $N$ TTS models with the softmax function over distances to each TTS centroid. 
The centroid is:
\begin{equation}
    c_j = \frac{1}{M-1}\sum_{m=1}^{M-1} \mathtt{x}_{j,m}
\end{equation}
Instead of the $||\cdot ||_2^2$ norm which was proposed on the original paper of the prototypical loss, they used instead a similarity metric with learnable scale and bias parameters:
\begin{equation}
    \mathcal{S}_{j,k} = w\cdot \cos(\mathtt{x}_{j,M},c_k) + b
\end{equation}
The final loss is given by this equation:
\begin{equation}
    \mathcal{L}_{ap} = -\frac{1}{N}\sum_{j=1}^N \log \frac{\exp{\mathcal{S}_{j,j}}}{\sum_{k=1}^N \exp{\mathcal{S}_{j,k}}}
\end{equation}

\section{Dataset}

The dataset used in this source tracing task is a version of the MLAADv5 dataset designed specifically for source tracing. The dataset consists of three distinct and mutually exclusive subsets, a training set, a development set, and a test set. 

The training subset features 11.100 audio samples across 24 distinct classes, each class representing the TTS system that created the specific sample, all from 8 different languages.

The development subset features 12.000 audio samples across 25 distinct audio classes and 21 languages. However, in the development set there are languages and classes that were not seen during the training phase. More specifically, about 3000 samples are from seen models and seen languages, 1800 samples from seen models but unseen languages, 2700 samples from unseen models but seen languages, and 4500 samples where both the model and the language are unseen.

The test subset is comprised of 33900 samples from 37 languages and 64 different models, making it, in theory at least, even more difficult for our models to perform adequately here. From these 33900 samples, 8700 are from seen model and seen language, 7500 are from seen model but unseen language, 4800 are from seen language but unseen model and 12900 are from unseen language and unseen model.

\section{Experimental Setup}

For the purposes of this research we utilized the repository\footnote{\url{https://github.com/clovaai/voxceleb_trainer}} the authors of \cite{chung20b_interspeech} released, which we altered in order to work for this project, while enhancing its model pool with AASIST with fine-tuned Wav2Vec2 embeddings.

\subsection{Inputs}

During training, as input, we extracted randomly a 2s segment from each audio file. The input embeddings for the ResNet model were 40-dimensional Mel filterbanks while for the SSL we used 1024-dimensional embeddings from the output of the Wav2Vec model. Instance normalization was performed according to the original recipe of \cite{ulyanov2016instance}. Neither Voice Activity Detection nor Data Augmentation was applied.

\subsection{Base Model}

We experimented with two established base architectures during training, the Thin-ResNet model and the AASIST model with fine-tuned Wav2Vec embeddings. Thin-ResNet-34 is made up of 34 layers with one-quarter of the channels in each residual block. The total parameter count comes close to 1.4M parameters, with slight variations due to the dimensionality of the output embeddings. Self-Attentive Pooling (SAP) \cite{cai2018exploring} was also used to aggregate frame-level features into utterance-level representations. The SSL model consists of a pre-trained self-supervised Wav2Vec2 model\footnote{facebook/wav2vec2-xls-r-300m} for audio feature extraction, which is fine-tuned during training, and an AASIST model, exactly as described in \cite{jung2022aasist}, raising the total parameter count of the model to about 315M.

\subsection{Output Embeddings}

In all our experiments, we tested four different output embedding sizes: 10, 50, 200, and 512. In speaker recognition tasks, the optimal embedding size is typically well below the number of training speakers. However, since our dataset contains relatively few TTS systems, we experimented with both smaller and larger embedding dimensions.

\subsection{Implementation Details}

All our experiments were conducted on the Google Colab Pro+ platform and more specifically on runtimes with the A100 GPU and 83.5GB of VRAM. 
The batch size for the ResNet experiments with the random sampler was 128 and 12 for the ResNet experiments with the balanced sampler. For the SSL model the batch size was instead 32 for the random sampler experiments as 128 would not fit the GPU while fine-tuning the Wav2Vec2 model.
Each ResNet model was trained for 300 epochs, while each SSL model was trained for only 100 epochs. The ResNet models' trained for about 2 hours, while the SSL models' trained for about 4 hours. All of our experiments were run with a Cosine Annealing LR scheduler with linear warm-up for 10 epochs up to 1e-4. For the AMSoftmax and AAMSoftmax losses we used $m=0.3$ and $s=30$. 

Due to limited resources, we were not able to run the full extent of our experiments on the SSL model. We limited our loss selections, experimenting only with the AMSoftmax and the AAMSoftmax, while maintaining the full range of experimentation for the sampler and output embeddings. 

\subsection{Evaluation}

During training we evaluated the model at regular intervals (25 epochs for the ResNet model and 10 epochs for the SSL model) using the development dataset to extract a development EER score and assess performance. During inference, for each audio embedding, we computed its cosine similarity with every other audio sample in the set using the corresponding embeddings. These similarities were then used to calculate the EER.

To better understand the model's performance, we also performed Linear Probing \cite{NEURIPS2023_f249db9a} on the output embeddings of our best ResNet model. This involved training a single classification head on the development data output embeddings. The classification head was trained for 50 epochs with a learning rate of 0.1, a standard SGD optimizer, and the cross-entropy loss function. 
To address class imbalance, we applied undersampling to the output embeddings of three overrepresented classes: \textit{tts\_models/multilingual/multi-dataset/bark} (2400 samples), \textit{tts\_models/multilingual/multi-dataset/xtts\_v2} (1500 samples) and \textit{griffin\_lim} (1500 samples). After undersampling the dataset was reduced to 7500 samples with 300 samples per model class. We used 80\% of this balanced subset (6000 samples) to train the classification head, while the remaining 1500 embeddings were used to generate a confusion matrix. This allowed us to evaluate whether the original model’s embeddings were sufficiently separable for a single classification head to achieve reliable classification.

We performed this probing on the development set rather than the test set due to the number of classes. The development set contains 25 distinct classes—already sufficient to present a meaningful confusion matrix—whereas the test set has 64 classes, making it difficult to visualize effectively.

\section{Experimental Results}
The results for both the ResNet and the AASIST models can be found in Table \ref{tab:ResNet_results}.
The ResNet model exhibited mixed performance when trained with different classification loss functions. Notably, the softmax loss with a random sampler struggled to differentiate between audio samples across most output embedding sizes, with the exception of the 512-dimensional embeddings, where performance on both the development and test subsets was above average. In contrast, other configurations significantly outperformed this softmax-based approach. In particular, the AAMSoftmax loss with a random sampler consistently demonstrated strong performance across all embedding sizes, achieving one of the best development set results with the 50-dimensional embeddings.

\begin{table*}[h!]
    \caption{EER (\%) on the development and test sets attained by ResNet and AASIST models. R stands for experiments with a Random Sampler and B for experiments with Balanced Sampler.}
    \label{tab:ResNet_results}
    \centering
    \begin{tabular}{|r|r!{\vrule}c|cccc|cccc|}
    \hline
    \multirow{2}{*}{Model}&\multirow{2}{*}{Loss} & \multirow{2}{*}{Sampler}& \multicolumn{4}{c|}{EER(\%) - development set} & \multicolumn{4}{c|}{EER(\%) - test set}\\
    \cline{4-11}
    &  & & 10 & 50 & 200 & 512 & 10 & 50 & 200 & 512\\
    \hline
    \hline
    \multirow{8}{*}{\textbf{ResNet-34}} &\multirow{2}{*}{\textbf{Softmax}} & R & 44.14 & 44.18 & 47.29 & 7.75 & 47.98 & 48.90 & 49.81 & 8.76\\
     & & B & 8.09 & 7.70 & 7.57 & 8.37& 8.42 & 8.56 & 9.03 & 9.56 \\
     \cline{2-11}
     & \multirow{2}{*}{\textbf{AMSoftmax}} & R & 11.54 & 11.71 & 10.82 & 9.73& 13.98 & 13.71 & 11.77  & 11.93\\
     && B & 13.72 & 12.28 & 14.21 & 14.44& 16.39 & 13.80 & 16.07 & 16.63\\
     \cline{2-11}
     &\multirow{2}{*}{\textbf{AAMSoftmax}} & R & 6.64 & 5.57 & 6.05 & 6.40& 5.29 & \textbf{4.63} & 5.71 & 5.08\\
     && B & 11.35 & 9.47 & 10.14 & 12.06& 9.44 & 8.55 & 8.88 & 11.96\\
     \cline{2-11}
     &\textbf{AngularProto}& B & 6.57 & 6.08 & 5.46 & 6.12& 7.04 & 7.06 & 5.95 & 5.74\\
     \cline{2-11}
     &\textbf{GE2E} & B & 4.87 & \textbf{4.57} & 5.07 & 4.70& 5.42 & 5.03 & 4.94 & 5.17\\
     \hline
     \multirow{4}{*}{\textbf{AASIST}} & \multirow{2}{*}{\textbf{AMSoftmax}} & R & 5.22 & 5.43 & 5.59 & 6.1 & 6.04 & 5.76 & 5.53 & 8.76 \\
     & & B & 5.45 & 5.18 & 5.31 & 6.53 & 6.03 & 6.13 & 6.59 & 6.90 \\
     \cline{2-11}
     & \multirow{2}{*}{\textbf{AAMSoftmax}} & R & 5.07 & \textbf{4.49} & 5.01 & 5.23 &\textbf{4.77}&5.36& 5.53 &5.71\\ 
     &  & B & 5.61 & 5.47 & 6.35 & 5.78 &6.28&5.82& 5.99 &5.33\\
     \hline
    \end{tabular}
\end{table*}

The metric learning methods perform quite remarkably compared to most classification loss functions. Notably, GE2E with 50-dim. output embeddings is the only method applied to the ResNet model to give a sub-5\% development EER score, with 3 out of the 4 results being less than 5\%, and the only one over 5\% being a 5.07\% with the 200-dim. embeddings. 

In the test set, the results continue the same narrative. GE2E is again very competitive, with all four embedding dimensionalities being below 5.5\% and especially the 200-dim embedding having a 4.94\% EER. However, the best test result came from the AAMSoftmax loss with 50-dim embeddings and a Random Sampler. This particular model managed a 4.63\% EER on the test set, 0.31 percentage points (pp) below the best test EER of any ResNet model, which is surprising because, on the development set, this configuration was the only one below 6\% EER.

The AASIST model was better than any ResNet model on the development set with its best development EER score being a 4.49\% which is just 0.08 pp below the best ResNet development EER score. This score came from the AAMSoftmax loss paired with a Random Sampler and 50-dim. embeddings. A surprising result came from the evaluations on the test set as the best EER score did not come from the model that gave the best development EER, it came from the same loss function and Sampler, but with 10-dim. embeddings instead of 50-dim. The result was a 4.77\% EER which is 0.14 pp larger than the best test EER of the ResNet model.

AASIST models generally outperform ResNet on the development set but exhibit weaker generalization, as their test EER is usually higher than their development EER. Among the eight AASIST models with AAMSoftmax loss only three performed better on the test set, whereas all ResNet models with AAMSoftmax showed improvement. This may be due to Wav2Vec2 embeddings learning specific features that may do not always aid with unseen classes, unlike Mel filterbank features, which remain consistent across seen and unseen classes. While not true for all ResNet configurations (as shown in Table \ref{tab:ResNet_results}), this trend is evident in most AASIST models. It is also evident that most models do not benefit from larger output embedding sizes and keeping the dimensionality between 10 and 50 seems to work just fine when compared with 200 and 512.

\begin{figure}[h!]
    \centering
    \includegraphics[width=0.9\linewidth]{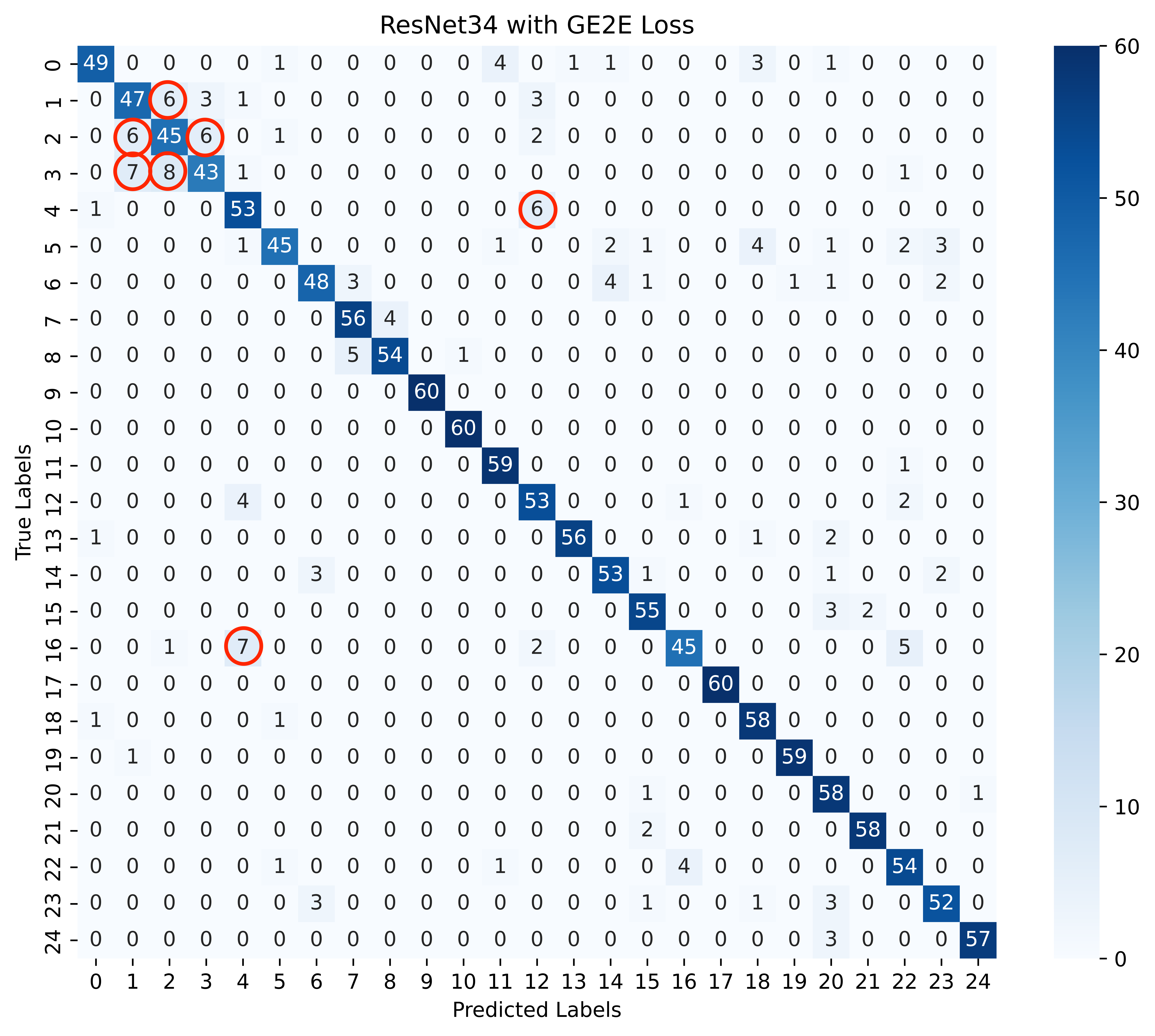}
    \caption{Confusion Matrix of the 50-dim output embeddings of the ResNet-34 model trained with the GE2E loss. The most notable misclassifications are marked with a red circle.}
    \label{fig:conf_mat}
\end{figure}

The confusion matrix in Fig. \ref{fig:conf_mat} shows that a single classification head generally distinguishes between classes well, with few misclassifications. However, classes 1, 2, and 3\textemdash three variations of the \textit{facebook/mms-tts} model trained in different languages\textemdash are exceptions, likely due to their shared architecture. Our ResNet model was trained on this architecture in German. Other notable misclassifications include seven \textit{tts\_models/hr/cv/vits} samples misclassified as \textit{griffin\_lim} and six \textit{griffin\_lim} samples misclassified as \textit{tts\_models/es/css10/vits}. Since the ResNet model was trained with these architectures in different languages, this may not be coincidental and could be explored further. Despite these misclassifications, the ResNet model produced embeddings distinct enough for a linear classifier to separate, even for unseen classes. This is a strong indication that this speaker recognition pipeline can be extended to source tracing and be a competitive solution.

\begin{figure}
    \centering
    \includegraphics[width=0.9\linewidth]{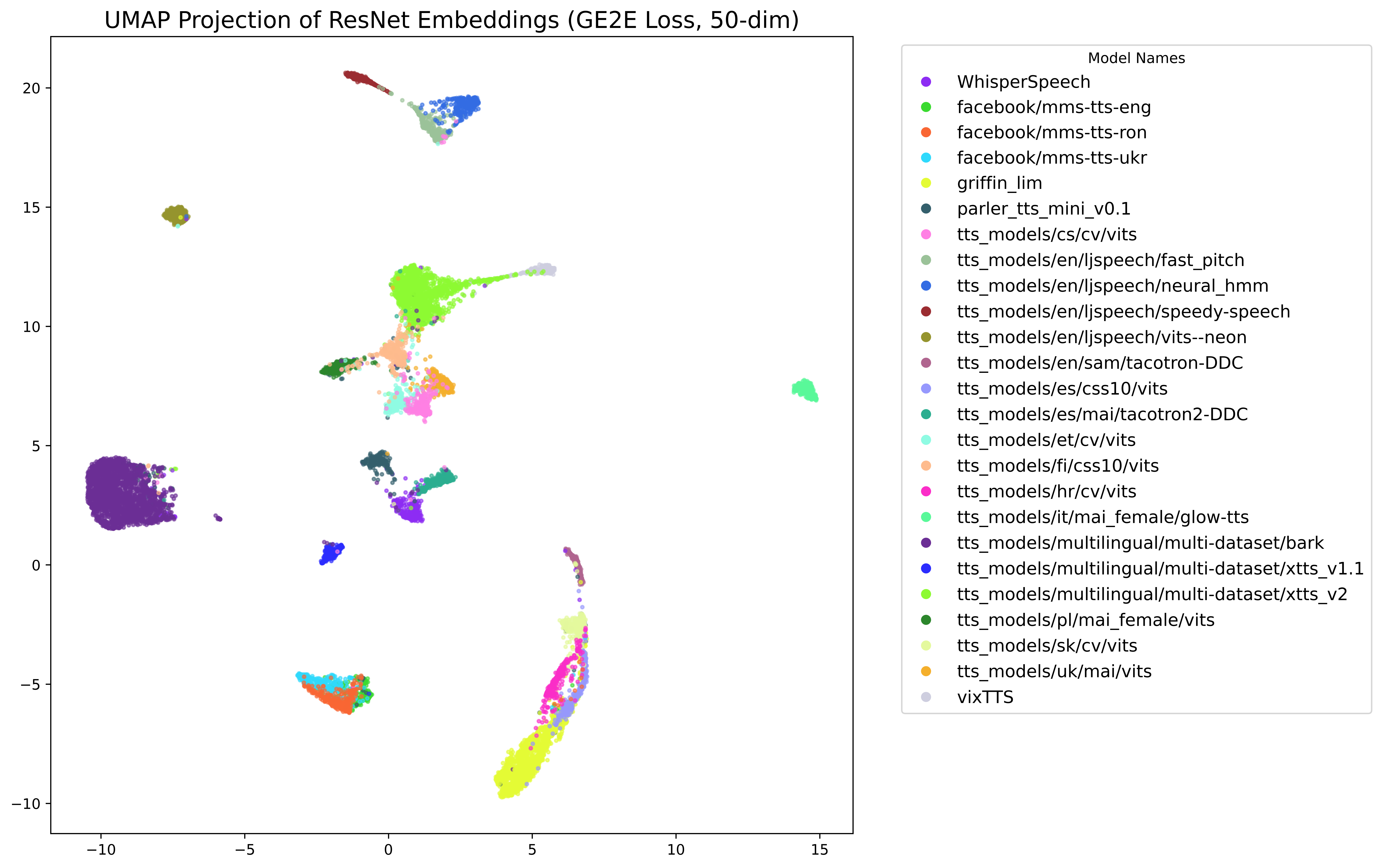}
    \caption{UMAP projection of the embeddings from the development set extracted with ResNet (GE2E loss, 50-dim).}
    \label{fig:umap}
\end{figure}

In Fig. \ref{fig:umap}, we visualize the embeddings of the same model using UMAP \cite{McInnes2018UMAPUM}. Some of the classes are well separated from all other classes and thus can be separated better by the model, like the \textit{bark} embeddings on the left of the diagram. In contrast, mms-tts variants at the bottom-left exhibit overlapping embeddings, forming a single mixed cluster—consistent with their frequent misclassification in the confusion matrix (Fig. \ref{fig:conf_mat}). Overlaps are also evident in the center cluster, mainly by vits models, and the lower-right cluster, where \textit{griffin\_lim} embeddings are mixed with those of \textit{tts\_models/hr/cv/vits} and \textit{tts\_models/es/css10/vits}. These findings are in line with the observations we previously made about these classes from the examination of the confusion matrix (Fig.\ref{fig:conf_mat}).

\section{Conclusions}

We present an approach to the synthetic speech source tracing task that is largely inspired by the speaker recognition pipeline. Our experiments show that a simple ResNet model with a GE2E loss and 50-dim embeddings is a viable solution to discern between the source model of a spoofed sample. We also provide evidence that the size of the embeddings can be as small as 10 with only slight degradation. Further experimentation may include other architectures and training algorithms, as well as experiments with replayed TTS. 

\section{Acknowledgements}

This work has been partially supported by project MIS 5154714 of the National Recovery and Resilience Plan Greece 2.0 funded by the European Union under the NextGenerationEU Program.

\bibliographystyle{IEEEtran}
\bibliography{mybib}

\end{document}